\documentclass[aps,preprint,epsfig,rotate]{revtex4}
\usepackage{graphicx}
\usepackage{bm}
\usepackage{epsfig}

\input epsf
\begin{document}
\title{\bf On the classical theory of molecular optical activity}

 \author{Alexei M. Frolov}
 \email[E--mail address: ]{afrolov@uwo.ca}

 \author{David M. Wardlaw}
 \email[E--mail address: ]{dwardlaw@uwo.ca}

\affiliation{Department of Chemistry\\
 University of Western Ontario, London, Ontario N6H 5B7, Canada}

\date{\today}

\begin{abstract}

The basic principles of classical and semi-classical theories of molecular
optical activity are discussed. These theories are valid for dilute
solutions of optically active organic molecules. It is shown that all
phenomena known in the classical theory of molecular optical activity can be
described with the use of one pseudo-scalar which is a uniform function of
the incident light frequency $\omega$. The relation between optical rotation
and circular dichroism is derived from the basic Kramers-Kronig relations.
In our discussion of the general theory of molecular optical activity we
introduce the tensor of molecular optical activity. It is shown that to
evaluate the optical rotation and circular dichroism at arbitrary
frequencies one needs to know only nine (3 + 6) molecular tensors. The
quantum (or semi-classical) theory of molecular optical activity is also
briefly discussed. We also raise the possibility of measuring the optical
rotation and circular dichroism at wavelengths which correspond to the
vacuum ultraviolet region, i.e. for $\lambda \le 150$ $nm$.

PACS number(s): 33.55.+b and 33.20.Ni
\end{abstract}
\maketitle

\section{Introduction}\label{intro}

In this study we discuss the classical theory of molecular optical activity.
This theory was originally developed for solutions of various optically
active organic molecules (see, e.g., \cite{Mason}, \cite{Rober} and
references therein). Our analysis begins with the classical theory of
optical activity based on the Maxwell's equations for electromagnetic
fields. Any optical active substance is described in this theory with the
use of a few phenomenological parameters. The main goal of the classical
theory of optical activity is to derive some useful relations between these
parameters in various cases. In general, these parameters also depend upon
frequencies and relations between such parameters take different forms for
different frequencies. We also consider semi-classical theory of optical
activity of molecules which was originally developed by Rosenfeld in
\cite{Rose}. In this theory all molecules are considered as quantum systems,
while radiation is considered classically. This old theory is still widely
used, since it produces a very good agreement with many experimental
results. In particular, the semi-classical theory of optical activity can be
used at short and very short wavelengths, e.g., for wavelengths which
correspond to vacuum ultraviolet. On the other hand, it is clear that the
complete theory of optical activity can be based only on quantum mechanics
of molecules and quantum theory of radiation.

This work has the following structure. Basics of the classical theory of
molecular optical activity in dilute solutions of organic substances can be
found in the next Sections. Here we introduce the optical rotatory parameter
$\beta$. The four Stokes parameters are defined in Section III. These
parameters are very convenient to describe quasi-monochromatic light. The
phenomenon of circular dichroism is described in Section IV. It appears that
the two fundamental $\omega-$dependent functions (optical rotation and
circular dichroism) which can be defined for an arbitrary optically active
solution can be written in the form of one complex function. The well known
Kramers-Kronig relation between the real and imaginary parts of this
functions must always be obeyed. For limited intervals of frequencies this
produces a very useful relation which allows to determine, e.g., the
circular dichroism by using the known expressions for optical activity.
Tensor of molecular optical activity is explicitly defined in Section V.
The formulas obtained in this Section are very useful in applications, since
they allow to express the optical activity by using only the two basic
molecular properties (the electric dipole and magnetic dipole momenta).
Concluding remarks can be found in the last Section.

\section{Classical theory of molecular optical activity}

Let us briefly discuss the classical theory of optical activity. In
classical theory the optical activity always denotes the ability of the
material under study to rotate the plane of polarization of the left- and
right-circularly polarized light. Currently, the study of optical activity
also includes optical rotation at different wavelengths, circular dichroism,
and differential scattering of left- and right-circularly polarized light.
All these phenomena are manifestations of natural optical activity which is
a characteristic of chiral molecules (in contrast with achiral or non-active
molecules). Note that there are also various phenomena which correspond to
so-called induced optical activitivies. In such cases the achiral molecules
can show some optical activity, if, e.g., they are placed in a relatively
strong electric and/or magnetic fields. In this study we restrict ourselves
to the analysis of the natural optical activity only.

In general, the optical activity is uniformly related to the spatial
dispersion, i.e. to the non-local relation between the electric induction
${\bf D}$ and electric field ${\bf E}$. For the Cartesian components of
these vectors we can write \cite{LLC}
\begin{equation}
 D_i({\bf r},t) = E_i({\bf r},t) + \int^{\infty}_0 d\tau \int d^3{\bf
 r}_1 F_{ij}(\tau, {\bf r}, {\bf r}_1) E_j(t-\tau,{\bf r}_1) \label{e1.1}
\end{equation}
where $F_{ik}(\tau, {\bf r}, {\bf r}_1) = F_{ki}(\tau, {\bf r}_1, {\bf r})$
is the kernel of integral operator. For monochromatic field components
${\bf E}({\bf r},t) = {\bf E}({\bf r}) \exp(-\imath \omega t)$ and
Eq.(\ref{e1.1}) takes the form
\begin{equation}
 D_i({\bf r}) = E_i({\bf r}) + \int d^3{\bf r}_1 f_{ij}(\omega; {\bf r},
 {\bf r}_1) E_j({\bf r}_1) \label{e1.2}
\end{equation}
This equation with the kernel $f_{ij}(\omega; {\bf r}, {\bf r}_1)$ expresses
a non-local relation between ${\bf D}$ and ${\bf E}$ which is also called
spatial dispersion. In general, the kernel $f_{ij}(\omega; {\bf r}, {\bf
r}_1)$ in Eq.(\ref{e1.2}) rapidly decreases with interatomic distances. In
many cases such a kernel is very small already at distances $\approx$ 3 $a$,
where $a$ designates some average (or effective) atomic dimension. Briefly,
the relation, Eq.(\ref{e1.2}), is written in the form
\begin{equation}
 D_i({\bf r}) =  \int d^3{\bf r}_1 \cdot \epsilon_{ij}(\omega; {\bf r},
 {\bf r}_1) E_j({\bf r}_1)
\end{equation}
where $\epsilon_{ij}$ is some non-local operator (tensor) which also depends
upon the frequency $\omega$ (see below). The $3 \times 3$ tensor
$\hat{\epsilon} = \epsilon_{ij}$ is the dielectric tensor (also called
electric permittivity). From the transparency of the media it follows that
all nine matrix elements $\epsilon_{ij}$ are real and all three eigenvalues
of this tensor are positive.

In all studies of optical activity in organic compounds only infinite,
homogeneous media are considered, a convention adopted in this work. Unless
otherwise specified, the absorbtion of radiation is assumed to be absent at
all frequencies considered below. In such cases the kernel in
Eq.(\ref{e1.2}) depends only on the difference ${\bf R} = {\bf r} - {\bf
r}_1$. The functions ${\bf D}$ and ${\bf E}$ in infinite, homogeneous media
can be expanded in a Fourier integral with the respect to Cartesian
coordinates as well as time. Finally, this allows one to obtain the
following relation between the corresponding Cartesian components of the
vectors ${\bf D}$ and ${\bf E}$
\begin{eqnarray}
 D_i({\bf k}) = \epsilon_{ij}(\omega; {\bf k}) E_j({\bf k}) =
 \Bigl[\delta_{ij} + \int^{\infty}_0 d\tau \int d^3{\bf R} f_{ij}(\tau,{\bf
 R}) \exp[\imath (\omega \tau - {\bf k} \cdot {\bf R})] \Bigr]
 E_j({\bf k}) \label{e1.3}
\end{eqnarray}
In other words, the dielectric tensor $\epsilon_{ik}(\omega; {\bf k})$ (also
called electric permittivity) takes the form
\begin{eqnarray}
 \epsilon_{ij}(\omega,{\bf k}) = \delta_{ij} + \int^{\infty}_0 d\tau \int
  d^3{\bf R} f_{ij}(\tau,{\bf R}) \exp[\imath (\omega \tau - {\bf k} \cdot
 {\bf R})]
\end{eqnarray}
As follows from this formula the dielectric tensor is a function of the
field frequency $\omega$ and wave vector ${\bf k}$. In general, the
dependence of the dielectric tensor $\epsilon_{ij}$ on $\omega$ is called
dispersion, while the analogous dependence upon the wave vector ${\bf k}$
represents the spatial dispersion. The spatial dispersion of
$\epsilon_{ij}({\bf k})$ is responsible for optical activity (see below).

In solutions of organic substances the optical activity corresponds to the
case of weak spatial dispersion, i.e. $k = \mid {\bf k} \mid$ is small. In
such cases the tensor $\varepsilon_{ij}(\omega,{\bf k})$ can be expanded in
powers of the wave vector ${\bf k}$, e.g.,
\begin{eqnarray}
 \epsilon_{ij}(\omega,{\bf k}) = \epsilon^{(0)}_{ij}(\omega) +
 \gamma_{ijl}(\omega) k_l + \beta_{ijlm}(\omega) k_l k_m +
 \alpha_{ijlmn}(\omega) k_l k_m k_n + \ldots \label{expan}
\end{eqnarray}
Such an expansion is valid, if the first term in Eq.(\ref{expan}), i.e.
$\epsilon^{(0)}_{ij}(\omega)$, has no zeros in a given range of frequencies
$\omega$. Since in this study we restrict ourselves to the consideration of
transparent (or slightly absorbing) solutions only, then we can neglect the
imaginary part of dielectric tensor $\epsilon^{(0)}_{ij}(\omega)$.

If these two conditions are obeyed, then for small ${\bf k}$ only the first
few terms in such an expansion are important. Let us restrict to the two
lowest order terms (the second of which is responsible for optical
activity), i.e. we can write
\begin{equation}
 \epsilon_{ij}(\omega,{\bf k}) = \epsilon^{(0)}_{ij}(\omega) +
 \gamma_{ijl}(\omega) k_l = \epsilon^{(0)}_{ij}(\omega) + \imath
 \frac{\omega}{c} \gamma_{ijl}(\omega) n_l \label{e7}
\end{equation}
where ${\bf n} = \frac{c}{\omega} {\bf k}$ and $\gamma_{ijl} n_l$ is
an antisymmetric tensor of the second rank (upon indexes $i$ and $j$). For
tensor $\gamma_{ikl}$ the antisymmetry means $\gamma_{ijl} = -\gamma_{jil}$.
If absorbtion of radiation is absent, then the tensor $\gamma_{ijl}$ is
real, i.e. $\gamma^{*}_{ijl} = \gamma_{ijl}$. These two conditions mean that
the $\gamma_{ijl} n_l$ tensor can be re-written into another form
$\gamma_{ijl} n_l = \frac{c}{\omega} e_{ijl} g_l$, where $e_{ijl}$ is the
complete antisymmetric tensor, while $g_l$ is the $l-$th component of the
axial giration vector ${\bf g}$. In tensor algebra this relation is called
the duality relation. In general, the giration vector ${\bf g}$ is a
function of the unit wave vector ${\bf n}$, i.e. $({\bf g})_i = g_{il} n_l$,
where $g_{il}$ is the pseudotensor of the second rank. In isotropic media
$g_{il} = \delta_{il} f$ and such a pseudotensor is reduced to a single
pseudoscalar $f$. In this case the $\gamma_{ijl}$ tensor is essentially
reduced to the complete antisymmetric tensor $e_{ijl}$.

In fact, for the tensor $\gamma_{ijl}$ one finds $\gamma_{ijl}(\omega) =
\frac{c}{\omega} e_{ijl} f(\omega)$, and therefore,
\begin{equation}
 {\bf D} = \epsilon^{(0)} {\bf E} + \imath f(\omega) ({\bf E}
 \times {\bf n}) \label{Main}
\end{equation}
Note again that this equation can be applied only in those cases when the
absolute value of $f(\omega)$ is much smaller than the minimal eigenvalue of
the tensor $\epsilon^{(0)}(\omega)$.

As is well known (see, e.g., \cite{LLC}) in an arbitrary dielectric media
we always have ${\bf D} \cdot {\bf n} = 0$. In this case from
Eq.(\ref{Main}) one also finds that ${\bf E} \cdot {\bf n} = 0$. For a
monochromatic wave we can write the Maxwell equations
\begin{equation}
 \frac{\omega}{c} {\bf H} = ({\bf k} \times {\bf E}) \; \; \; and \; \; \;
 \frac{\omega}{c} {\bf D} = ({\bf k} \times {\bf H}) \label{e8}
\end{equation}
It follows from here that ${\bf k} \perp {\bf D} \perp {\bf H}$ and also
that ${\bf E} \perp {\bf H}$. In three-dimensional space this means that the
three vectors ${\bf E}, {\bf D}$ and ${\bf k}$ are coplanar. This simplifies
drastically the following analysis of optical activity in isotropic media.

Consider now the energy transfer. In general, the direction of the energy
flux is given by Poynting vector ${\bf S} = \frac{c}{4 \pi} ({\bf E} \times
{\bf H})$. Now by using the unit vector ${\bf n}$ defined above (${\bf n} =
\frac{c}{\omega} {\bf k}$) we can write for the Poynting vector
\begin{equation}
 {\bf S} = \frac{c}{4 \pi} ({\bf E} \times {\bf H}) = \frac{c}{4 \pi} [{\bf
 n} E^2 - ({\bf E} \cdot {\bf n}) {\bf E}]
\end{equation}
The total energy flux through an element $dS$ of surface orthogonal to ${\bf
n}$ is
\begin{equation}
 dW = \frac{c}{4 \pi} [E^2 - ({\bf E} \cdot {\bf n})^2] d\Omega =
 \frac{c}{4 \pi} E^2 sin^3 \Theta d\Theta d\Phi
\end{equation}
where $\Theta$ is the angle between the vector ${\bf E}$ and outer normal to
this surface element $dS$, i.e. ${\bf n}$. Also, it follows from the two
equations of Eq.(\ref{e8}) that ${\bf D} = n^2 {\bf E} - {\bf n} ({\bf n}
\cdot {\bf E})$. On the other hand the basic relation between vectors ${\bf
D}$ and ${\bf E}$ is ${\bf D} = \hat{\epsilon} {\bf E}$, where
$\hat{\epsilon}$ is the dielectric tensor. From here one finds the following
equation written in Cartesian components
\begin{equation}
 (n^2 \delta_{ij} - n_i n_j - \epsilon_{ij}) E_j = 0 \label{fren}
\end{equation}
where $\epsilon_{ij}$ are the components of dielectric tensor.

Formally, this equation coincides with the corresponding eigenvalue equation
for the dielectric tensor $\epsilon_{ik}$. However, the eigenvalues of this
tensor are the functions of three spatial directions. By using some unitary
transformation, one can reduce Eq.(\ref{fren}) to the principal axes of the
tensor $\epsilon_{ij}$ which are also called the principal dielectric axes.
In fact, there are some advantages to writing Eq.(\ref{fren}) in the
principal dielectric axes. In this case it exactly coincides with Fresnel's
equation which is the main equation of crystal optics. In general,
Eq.(\ref{fren}) determines the wave-vector surface in the $n_x, n_y, n_z$
coordinates. Such surfaces depend upon three constant coefficients
$\epsilon_{x}, \epsilon_{y}, \epsilon_{z}$ (eigenvalues of the dielectric
tensor $\epsilon_{ij}$).

For homogeneous solutions the Fresnel's equation simplifies significantly,
since is these systems $\epsilon_{x} = \epsilon_{y} = \epsilon_{z}$. We want
to consider such a transition in the two following steps. First, consider
the case of two different eigenvalues $\epsilon_{x} = \epsilon_{y} =
\epsilon_{\perp}$ and $\epsilon_{z} = \epsilon_{\parallel}$ (these values of
parameters correspond to uniaxial crystals). In this case the Fresnel's
equation can be factorized to the form
\begin{equation}
 ( n^2 - \epsilon_{\perp} ) [ \epsilon_{\parallel} n^2_z + \epsilon_{\perp}
 ( n^2_x + n^2_y ) - \epsilon_{\perp} \epsilon_{\parallel} ] = 0
\end{equation}
where ${\bf n} = (n_x, n_y, n_z)$ is the direction of the light propagation.
In other words, an equation of the fourth order (upon $n$) is reduced to
the product of the two quadratic equations
\begin{eqnarray}
 && n^2 = n^2_x + n^2_y + n^2_z = \epsilon_{\perp} \label{e14} \\
 && \frac{n^2_z}{\epsilon_{\perp}} + \frac{n^2_x +
     n^2_y}{\epsilon_{\parallel}} = 1 \label{e15}
\end{eqnarray}
where the first equation is the equation of a sphere, while the second
equation determines an ellipsoid. The sphere represents the propagation of
ordinary waves. Such waves have the same refractive index $n =
\sqrt{\epsilon_{\perp}}$. The second equation represents the so-called
extraordinary waves which are directly related with the optical activity.
Let us consider the extraordinary waves in homogeneous solutions, or in
crystals of a cubic system. These two cases can be obtained as the
limit of Eq.(\ref{e15}) when $\epsilon_{\parallel} \rightarrow
\epsilon_{\perp}$. In reality, it can be written in the two different forms
$\epsilon_{\parallel} = \epsilon_{\perp} \pm \Delta$, where the positive
parameter $\Delta \rightarrow 0$. In such cases, Eq.(\ref{e14}) does not
change, while the second equation takes the form
\begin{equation}
 \frac{n^2_z}{\epsilon_{\perp}} + \frac{n^2_x +
  n^2_y}{\epsilon_{\perp} \pm \Delta} = 1 \; \; \; or \; \; \;
 \frac{n^2 cos^2\theta}{\epsilon_{\perp}} + \frac{n^2
 sin^2\theta}{\epsilon_{\perp} \pm \Delta} = 1
\end{equation}
where $n_x = n \cdot \sin\theta \cos\phi, n_y = n \cdot \sin\theta \sin\phi,
n_z = n \cdot \cos\theta$, where $\theta$ is the angle between the optical
axis and vector ${\bf n}$.

In homogeneous solutions the orientation of chiral molecules is random,
i.e. we have to replace the factors $cos^2\theta$ and $sin^2\theta$ in the
last equation by their mean values $\frac12$. This gives
\begin{equation}
 \frac{1}{\epsilon_{\perp}} + \frac{1}{\epsilon_{\perp} \pm \Delta} =
 \frac{2}{n^2}
\end{equation}
or
\begin{equation}
 \frac{2}{n^2} = \frac{\epsilon_{\perp} + \epsilon_{\perp} \pm
 \Delta}{\epsilon_{\perp} (\epsilon_{\perp} \pm \Delta)} \approx
 \frac{2 (\epsilon_{\perp} \pm \frac12 \Delta)}{(\epsilon_{\perp} \pm
 \frac12 \Delta)^2} = \frac{2}{(\epsilon_{\perp} \pm \frac12 \Delta)}
\end{equation}
From here one finds that $n^2 = \epsilon_{\perp} \pm \frac12 \Delta$, or in
other words, we have two different refractive indices $n^2_1 =
\epsilon_{\perp} + \frac12 \Delta$ and $n^2_2 = \epsilon_{\perp} - \frac12
\Delta$. This means that two different refracted wave are formed and,
formally, we have to consider the double refraction or birefringence.
However, the parameter $\Delta$ is small (in fact, very small) in comparison
with $n^2$. Therefore, the overall scale of such a birefringence is $\approx
\Delta$.

Result can be obtained in a slightly different way with the use of some
microscopic identities. Indeed, let us note that for homogeneous solutions
of chiral molecules $\epsilon_{ij} = \epsilon \cdot \delta_{ik} + \imath
\frac{c}{\omega} f(\omega) e_{ikl} n_l$, where $e_{ikl}$ is the complete
antisymmetric tensor. In this case we do not need to use the complete
Fresnel's equation to produce the same answer as above. The chiral activity
can be described with the use of only one numerical parameter $f(\omega)$
which is pseudoscalar. In fact, such a parameter can be introduced in a
slightly different way. Indeed, the Maxwell equations in the case of
homogeneous solutions take the form
\begin{eqnarray}
 {\bf D} = \epsilon {\bf E} - g \frac{\partial{\bf H}}{\partial t} \; \; \;
 , \; \; \; {\bf B} = \mu {\bf H} + g \frac{\partial{\bf E}}{\partial t}
\end{eqnarray}
where $g$ is a some constant, which depends upon $\omega$. By taking into
account polarization of media by the electric and magnetic field we can
write, e.g., for the ${\bf D}$ vector
\begin{eqnarray}
 {\bf D} = (1 + 4 \pi N \alpha) {\bf E} - 4 \pi N \frac{\beta}{c}
 \frac{\partial{\bf H}}{\partial t} \label{Ebas}
\end{eqnarray}
where $\alpha$ is the static polarizability, while $\beta$ is the so-called
optical rotatory parameter, or optical rotation, for short. As follows from
Eq.(\ref{Ebas}) the parameter $\beta$ is also a pseudoscalar. The parameter
$\beta$ determines the optical rotation, i.e. the rotation of the plane of
left- and right-circularly polarized light when it passes through an
optically active medium. Also, in this equation $N$ is the number of chiral
molecules per unit volume. It follows from the last two equations that
$\epsilon = 1 + 4 \pi N \alpha$ and $g = 4 \pi N \frac{\beta}{c}$. The
relation between factor $g$ and the indices of refraction for circularly
polarized light can also be obtained from Eq.(\ref{Ebas})
\begin{eqnarray}
 \chi_L = \sqrt{\epsilon} - 2 \pi \omega g \; \; \; and \; \; \;
 \chi_R = \sqrt{\epsilon} + 2 \pi \omega g
\end{eqnarray}
Now it is easy to find the overall rotation ($\delta$) when the light
propagates the distance $z$ in some chiral media
\begin{eqnarray}
 \delta = \frac{\pi z}{\lambda} (\chi_R - \chi_L) = \frac{4 \pi^2
 z}{\lambda^2} g = 4 \pi^2 \nu^2 z \cdot g = \frac{16 \pi^3 \nu^2 N
 z}{c} \cdot \beta \label{Ebas1}
\end{eqnarray}
Note that the optical rotatory parameter $\beta$ (as well as $\alpha$) which
follows from Eq.(\ref{Ebas}) can rigorously be determined only with the use
of the quantum mechanics of molecules. This will be our goal in the third
Section.

\section{Stokes parameters}

As follows from its definition any monochromatic wave has a certain
polarization. However, in actual optical experiments it is almost impossible
to create a beam of pure monochromatic waves, and usually we have to operate
with real light which contains frequencies distributed in a small interval
$\Delta \omega$ around the main frequency $\omega$. The means that the real
light is, in fact, a mixture of light quanta with different polarizations.
An arbitrary property of such a beam of light, e.g., the electric field
${\bf E}$, in the real light depends upon time. If the frequency
distribution $\Delta \omega$ around $\omega$ is narrow, then the ${\bf
E}(t)$ function can be represented in the form ${\bf E}(t) = {\bf E}_0(t)
exp(-\imath \omega t)$, where ${\bf E}_0(t)$ is a slowly varying function of
time $t$ which determines the polarization of the actual light. From the
last formula one can expect that such a polarization will be slowly changing
in time, i.e. we are dealing with the partially polarized light \cite{LLCE}.

In regular experiments we cannot observe the polarization properties of
electromagnetic waves directly. Instead, one measures the intensity
distribution of light when it passes through various physical bodies. This
means we are dealing with quadratic functions of the field. In other words,
in actual experiments we are measuring the components of the following
tensor $J_{\alpha\beta} = \overline{E_{0\alpha} E^{*}_{0\beta}}$, where
$E_{0\alpha}$ and $E^{*}_{0\beta}$ are the Cartesian components of the
slow varying ${\bf E}_0(t)$ vector. The line over the product of the two
complex vectors mean the value averaged in time. If all vectors are
represented in the form ${\bf E}(t) = {\bf E}_0(t) exp(-\imath \omega t)$,
then the $E_{0\alpha} E^{*}_{0\beta}$ product is the only value which has
non-zero time-average. Other similar combinations, i.e., $E^{*}_{0\alpha}
E^{*}_{0\beta}$ and $E_{0\alpha} E_{0\beta}$, contain rapidly oscillating
factors $exp(-2 \imath \omega t)$ which gives zero upon time averaging.

Since in any plane wave one finds ${\bf E} \perp {\bf n}$, where ${\bf n}$
is the direction of wave propagation, then the $J_{\alpha\beta}$ tensor has
only four components. Moreover, the $J_{\alpha\beta}$ tensor also contains
the total intensity of the wave $J = \sum_{\alpha} J_{\alpha\alpha} =
\overline{{\bf E} \cdot {\bf E}^{*}}$. This value has nothing to do with
with polarization of the wave and can be excluded by introducing the tensor
$\rho_{\alpha\beta} = \frac{J_{\alpha\beta}}{J}$. The tensor
$\rho_{\alpha\beta}$ has the unit trace and it is called the polarization
tensor. It can be shown that the polarization tensor is hermitian, i.e.
$\rho^{*}_{\alpha\beta} = \rho_{\beta\alpha}$. Now, we can introduce the
degree of polarization $P$ which is defined as
\begin{eqnarray}
 P = \sqrt{1 - 4 det(\rho_{\alpha\beta})} = \sqrt{1 - 4 \rho_{11} \rho_{22}
 + 4 \mid\rho_{12}\mid^2}
\end{eqnarray}
where $det(\rho_{\beta\alpha})$ is the determinant of the $2 \times 2$
matrix $\rho_{\beta\alpha}$.

An arbitrary hermitian $2 \times 2$ matrix can be represented in the
following form
\begin{eqnarray}
 \rho_{\alpha\beta} = \frac12 (\rho_{\alpha\beta} +
 \rho_{\beta\alpha}) + \frac12 (\rho_{\alpha\beta} -
 \rho_{\beta\alpha}) = S_{\alpha\beta} - \frac{\imath}{2}
 e_{\alpha\beta} A
\end{eqnarray}
where $S_{\alpha\beta}$ is the real symmetric $2 \times 2$ tensor. The
analogous non-symmetric $2 \times 2$ tensor in two-dimensional space is
reduced to the unit antisymmetric tensor $e_{12} = - e_{21}$ and
pseudoscalar $A$. The pseudoscalar $A$ is called the degree of circular
polarization. It is bounded between -1 and +1. The case of a light wave with
linear polarization corresponds to the $A = 0$ value. The waves with
circular polarization correspond to the values $A = +1$ (right-circular)
polarization and $A = -1$ (left-circular polarization).

An alternative expansion for an arbitrary hermitian $2 \times 2$ matrix is
performed with the use of three Pauli matrices $\hat{\sigma}_i$ ($i = x, y,
z$) (see, e.g., \cite{Dir}) and one unit matrix $\hat{I}$. It is written in
the form
\begin{equation}
 \rho_{\alpha\beta} = \frac12 \Bigl(\hat{I} + \xi_1 \hat{\sigma}_x +
 \xi_2 \hat{\sigma}_y + \xi_3 \hat{\sigma}_z\Bigr) =
 \frac12 \left(
 \begin{array}{cc}
  1 + \xi_3 & \xi_1 - \imath \xi_2 \\
  \xi_1 + \imath \xi_2 & 1 - \xi_3 \\
\end{array} \right) \nonumber
\end{equation}
The parameters $\xi_1, \xi_2$ and $\xi_3$ which appear in this formula are
the so-called Stokes parameters. In general, any intensity measurement may
be written as a linear combination of these three parameters and one
additional Stokes parameter $\xi_0$ which is the total intensity of the
scattered light. The determinant of the $\rho_{\beta\alpha}$ tensor is
\begin{eqnarray}
 det(\rho_{\alpha\beta}) = \frac14 (1 - \xi^2_1 - \xi^2_2 - \xi^2_3)
\end{eqnarray}
while the degree of polarization is $P = \sqrt{\xi^2_1 + \xi^2_2 +
\xi^2_3}$. The Stokes parameters $\xi_1$ and $\xi_3$ determine the degree of
the linear polarization, while the parameter $\xi_2$ shows the degree of
circular polarization. Note that the parameter $\xi_2$ coincides with the
pseudoscalar $A$ introduced above. From three Stokes parameters one can
construct the two scalars ($\xi_2$ and $\sqrt{\xi^2_1 + \xi^2_3}$) which are
invariants under Lorentz transformations. The three Stokes parameters also
have a number of other advantages in actual applications \cite{Ital}.

\section{Circular dichroism}

In all formulas above we have neglected the absorbtion of light during its
propagation in the dense media. In actual cases the absorbtion of light
always occurs. At some frequencies, e.g., in the vacuum ultraviolet region,
it plays a very important role and cannot be ignored even in the first
approximation. In reality the situation is even more complicated, since
light waves with different circular polarization are absorbed differently by
the media. This is called circular dichroism (CD). Such a differential
absorbtion of light with left- and right-circular polarizations can directly
(and substantially) affect the observed optical activity. It appears that
optical rotation and differential absorbtion of light with different
circular polarizations can be considered as the two manifestations of one
phenomenon.

In general, a detailed study of circular dichroism at different frequencies
allows one to develop a new approach for analysis of organic substances.
In this Section we want to discuss the modification which is required in all
formulas presented above. The absorbtion of light is described by
introducing an imaginary part into the permittivity tensor
$\epsilon_{ij}(\omega,{\bf r})$, or in other words, by considering static
polarizability $\alpha$ as a complex value. However, we are not interested
here in the total light absorbtion. Our interest is related to a very
specific difference between absorbtion of light with left- and
right-circular polarizations. It is clear that a complex static
polarizability $\alpha$, Eq.(\ref{Ebas}), cannot describe such differences.
As follows from Eq.(\ref{Ebas}) this goal can be achieved by considering the
optical rotatory parameter $\beta$ as a complex value.

In these cases the parameter $\beta$ is represented as the sum of its real
and imaginary parts, i.e. $\beta = \beta_1 + \imath \beta_2$, where $\beta_1$
and $\beta_2$ are two functions of the frequency $\omega$. These two
functions, however, are not completely independent, since there are two
additional connections between them which follow from the Kramers-Kronig
relations. This follows from the fact that $\beta(\omega)$ is the response
function \cite{LLC}, \cite{Jack} which is an analytical function in the
upper half $\omega$ plane (for now, we consider the frequency $\omega$ as a
complex variable) (see, e.g., \cite{Jack}). This allows us to use Cauchy's
theorem for $\beta(\omega)$:
\begin{equation}
 \beta(z) = \frac{1}{2 \pi \imath} \oint_C \frac{\beta(\omega^{\prime})
 d\omega^{\prime}}{\omega^{\prime} - z}
\end{equation}
The contour $C$ can be chosen to consist of the real frequency axis $\omega$
and a great semicircle at infinity in the upper half plane. The function
$\beta(\omega)$ vanishes rapidly at infinity, i.e. there is no contribution
to the integral from the great semicircle. Finally, Cauchy's integral is
written in the form
\begin{equation}
 \beta(z) = \frac{1}{2 \pi \imath} \int_{-\infty}^{+\infty}
 \frac{\beta(\omega^{\prime}) d\omega^{\prime}}{\omega^{\prime} - z}
\end{equation}
where $z$ now is any point in the upper $\omega-$half plane and the integral
is taken over the real axis. In fact, we want to place the point $z$ at the
real axis. This can be done by approaching the real axis from above, i.e.
by representing the complex variable $z$ in the form $z = \omega + \imath
\varepsilon$. This gives
\begin{equation}
 \beta(\omega) = \frac{1}{2 \pi \imath} \int_{-\infty}^{+\infty}
 \frac{\beta(\omega^{\prime}) d\omega^{\prime}}{\omega^{\prime} - \omega
 - \imath \varepsilon} \label{KK1}
\end{equation}

The denominator in the last formula can be written in the form \cite{Jack}
\begin{equation}
 \frac{1}{\omega^{\prime} - \omega - \imath \varepsilon} =
 P \Bigl(\frac{1}{\omega^{\prime} - \omega}\Bigr) + \pi \imath
 \delta(\omega^{\prime} - \omega)
\end{equation}
where the symbol $P$ means the principal part, while $\delta(x)$ designates
the Dirac delta-function. Now Eq.(\ref{KK1}) takes the form
\begin{equation}
 \beta(\omega) = \frac{1}{\pi \imath} P \int_{-\infty}^{+\infty}
 \frac{\beta(\omega^{\prime}) d\omega^{\prime}}{\omega^{\prime} - \omega}
 \label{KK2}
\end{equation}
By separating here the real and imaginary parts one finds
\begin{eqnarray}
 Re \beta(\omega) = \frac{1}{\pi} P \int_{-\infty}^{+\infty}
 \frac{Im \beta(\omega^{\prime}) d\omega^{\prime}}{\omega^{\prime} -
 \omega} \label{KK3} \\
 Im \beta(\omega) = -\frac{1}{\pi} P \int_{-\infty}^{+\infty}
 \frac{Re \beta(\omega^{\prime}) d\omega^{\prime}}{\omega^{\prime} -
 \omega} \nonumber
\end{eqnarray}
This is the most general Kramers-Kronig relations written for the
optical rotatory parameter $\beta$. In general, it can be shown that the
$Re \beta(\omega)$ is an even function in $\omega$, while $Im
\beta(\omega)$ is odd. This allows one to transform the last two integrals
in Eq.(\ref{KK3}) to the integrals taken over positive frequencies only,
i.e.
\begin{eqnarray}
 Re \beta(\omega) = \frac{2}{\pi} P \int_0^{+\infty}
 \frac{\omega^{\prime} [Im \beta(\omega^{\prime})]
 d\omega^{\prime}}{(\omega^{\prime})^2 - \omega^2} \label{KK4} \\
 Im \beta(\omega) = -\frac{2 \omega}{\pi} P \int_0^{+\infty}
 \frac{[Re \beta(\omega^{\prime})] d\omega^{\prime}}{(\omega^{\prime})^2 -
 \omega^2} \nonumber
\end{eqnarray}
These formulas can be used in actual applications which include the
optical rotatory parameter $\beta$. By using Eq.(\ref{Ebas1}) we can
re-write these formulas for the corresponding angles $\delta = \theta +
\imath \kappa$
\begin{eqnarray}
 \theta(\nu) = \frac{2}{\pi} P \int_0^{+\infty}
 \frac{\nu^{\prime} \kappa(\nu^{\prime})
 d\nu^{\prime}}{(\nu^{\prime})^2 - \nu^2} \label{KK5} \\
 \kappa(\nu) = -\frac{2 \nu}{\pi} P \int_0^{+\infty}
 \frac{\theta(\nu^{\prime}) d\nu^{\prime}}{(\nu^{\prime})^2 -
 \nu^2} \nonumber
\end{eqnarray}
where we have also introduced the linear frequency $\nu = \frac{\omega}{2
\pi}$ ($\omega$ is called the circular frequency). The importance of the
linear frequencies $\nu$ follows from the fact that these values are usually
used in actual experiments. In general, Eq.(\ref{KK5}) represents the
explicit relation between the actual optical rotation (angle $\theta$) and
circular dichroism (angle $\kappa$). As follows from Eq.(\ref{KK5}) the
optical rotation known for all frequencies allows one to determine the
circular dichroism at each frequency \cite{Djer}. In reality, however, one
finds a number of restrictions which exist in the solution of this problem.
Most of such restrictions follow from the fact that optical rotations in the
VUV region ($\lambda \le$ 150 $nm$) are not known even approximately. On the
other hand, it is clear that for each molecule the VUV area of wavelengths
contains many resonance lines which are crucially important to describe the
absorbtion of radiation. If we ignore the VUV region of wavelengths, then
we can restore the circular dichroism at all frequencies only approximately
(in reality, very approximately). For some limited areas of wavelengths,
however, such a reconstruction can be quite accurate and complete. Usually,
these areas of wavelengths are located far from the VUV region.

Note also, that the experimental knowledge of the $\theta(\nu)$ and
$\kappa(\nu)$ values for large number of different frequencies $\nu_1,
\nu_2, \ldots, \nu_n$ is used to detect uniformly the corresponding organic
substance. Formally, such an identification allows one to solve many
problems of quantative and qualitative analysis of the mixtures of chiral
organic substances.

\section{Tensor of molecular optical activity. Rotation power.}

In the middle of 1930's Placzek shown \cite{Pla} that a significant number
of effects related to the interaction between atom(s) and electromagnetic
field can be described with the use of only one tensor, later known as the
tensor of light scattering. In particular, the differential scattering
cross-section of light by an atom (or any other electron containing system)
can be written in the form
\begin{eqnarray}
 d\sigma = \frac{\omega (\omega + \omega_{12})^3}{\hbar^2 c^4} \mid
 (C_{ik})_{21} ({\bf e}^{\prime}_i)^{*} {\bf e}_k \mid^2 do^{\prime}
 \label{eq35}
\end{eqnarray}
where $(C_{ik})_{21}$ is the 3 $\times$ 3 tensor of light scattering, while
${\bf e}_i$ and ${\bf e}_k$ are the polarization vectors of the incident and
final photons. The integration in Eq.(\ref{eq35}) is performed over the
angular variables of the final photon which is designated by the superscript
${}^{\prime}$. Here and everywhere below we shall assume that the angular
volume element $do^{\prime}$ has the form $do^{\prime} = sin\theta^{\prime}
d\theta^{\prime} d\phi^{\prime}$. The explicit expression for the light
scattering tensor $(C_{ik})_{21}$ is \cite{LLQE}
\begin{eqnarray}
 (C_{ik})_{21} = \sum_{n} \Bigl[ \frac{(d_i)_{2n} (d_k)_{n1}}{\omega_{n1} -
 \omega -\imath 0} + \frac{(d_k)_{2n} (d_i)_{n1}}{\omega_{n1} +
 \omega^{\prime} -\imath 0} \Bigr] \label{eq36}
\end{eqnarray}
where $\omega^{\prime} = \omega + \omega_{12}$, while $d_i$ and $d_k$ are
the corresponding components of the vector of the dipole moment ${\bf d}$.
Note that the differential cross-section $d \sigma$, Eq.(\ref{eq35}),
corresponds to the lowest order approximation upon the fine structure
constant $\alpha \approx \frac{1}{137}$ and contains only the electric
dipole-dipole interaction. The method developed by Placzek \cite{Pla} was
based on eqalier studies by Kramers and Heisenberg \cite{Kram} and Dirac
\cite{Dirac}.

The Placzek approach for atoms suggests attempting to derive an analogous
method for molecules which would describe their optical activity. In this
Section this problem is considered in detail and it is shown that in the
lowest order approximation can be described by the tensor $(C_{ik})_{21}$ of
light scattering and by the four (or two in some cases) new tensors. These
tensors are called the tensors of (molecular) optical activity. Note that in
many actual cases the four/two tensors of optical activity are reduced to
one tensor only. To produce the closed analytical expressions for these
tensors below we shall assume that the electromagnetic field is represented
as a combination of plane waves. Each of these plane waves has its own
frequency $\omega$ and polarization which is represented by the vector ${\bf
e}$. The wave functions of the incident and final photons can be taken in
the form (see, e.g., \cite{LLQE})
\begin{eqnarray}
 {\bf A}_{{\bf e} \omega} = \sqrt{\frac{2 \pi}{\omega}} exp(-\imath \omega
 t + \imath {\bf k} \cdot {\bf r}) {\bf e}  \; \; \; , \; \; \;
 {\bf A}_{{\bf e}^{\prime} \omega^{\prime}} = \sqrt{\frac{2
 \pi}{\omega^{\prime}}} exp(-\imath \omega^{\prime} t + \imath {\bf
 k}^{\prime} \cdot {\bf r}^{\prime}) {\bf e}^{\prime} \; \; \; ,
\end{eqnarray}
where $\omega$ and $\omega^{\prime}$ are the corresponding frequencies,
while vectors ${\bf e}$ and ${\bf e}^{\prime}$ represent the polarization of
the incident and final photons, respectively. Below, we shall consider the
plane waves in the transverse (or radiation) gauge, where $div {\bf A} = 0$.
In this gauge one finds ${\bf k} \cdot {\bf e} = 0$ and ${\bf k}^{\prime}
\cdot {\bf e}^{\prime} = 0$. Note that in calculations for the final photon
we need to use the wave function which is conjugate to its wave function,
i.e. ${\bf A}^{*}_{{\bf e}^{\prime} \omega^{\prime}}$. As follows from these
equations the electric ${\bf E}$ and magnetic ${\bf H}$ fields are
\begin{eqnarray}
 {\bf E}_{{\bf e} \omega} = - \frac{\partial}{\partial t} {\bf A}_{{\bf e}
 \omega} = -\imath \sqrt{2 \pi \omega} {\bf e} exp(-\imath \omega t +
 \imath {\bf k} \cdot {\bf r}) \\
 {\bf H}_{{\bf e} \omega} = curl {\bf A}_{{\bf e} \omega} = \imath
 \sqrt{\frac{2 \pi}{\omega}} ({\bf k} \times {\bf e}) exp(-\imath \omega t
 + \imath {\bf k} \cdot {\bf r}) \nonumber
\end{eqnarray}
By introducing the unit vector ${\bf n} = \frac{{\bf k}}{\omega}$ we can
re-write the last equation in the form
\begin{equation}
 {\bf H}_{{\bf e} \omega} = \imath \sqrt{2 \pi \omega} ({\bf n} \times
 {\bf e}) exp(-\imath \omega t + \imath {\bf k} \cdot {\bf r})
\end{equation}
Analogous expressions can be obtained for the ${\bf E}_{{\bf e}^{\prime}
\omega^{\prime}}$ and ${\bf H}_{{\bf e}^{\prime} \omega^{\prime}}$ fields
which are related with the ${\bf A}^{*}_{{\bf e}^{\prime} \omega^{\prime}}$
wave function.

From these equations one finds the following expressions for the electric
dipole and magnetic dipole interactions. In fact, for each of the (${\bf e},
\omega$)-components of the ${\bf E}$ and ${\bf H}$ vectors we have
\begin{equation}
 V^{e}_{{\bf e} \omega} = -{\bf d} \cdot {\bf E}_{{\bf e} \omega} =
 \imath \sqrt{2 \pi \omega} ({\bf d} \cdot {\bf e}) exp(-\imath \omega t
 + \imath {\bf k} \cdot {\bf r}) \label{eq40}
\end{equation}
and
\begin{equation}
 V^{m}_{{\bf e} \omega} = -{\bf m} \cdot {\bf H}_{{\bf e} \omega} =
 - \imath \sqrt{2 \pi \omega} [{\bf m} \cdot ({\bf n} \times {\bf e})]
 exp(-\imath \omega t + \imath {\bf k} \cdot {\bf r}) \label{eq41}
\end{equation}
where ${\bf d}$ and ${\bf m}$ are the vectors of the electric and magnetic
dipole moments, respectively. In the lowest order approximation the
one-photon matrix elements of the $V^{e}$ and $V^{m}$ interactions equal
zero identically. The first non-zero contribution can be found only in the
second order of perturbation theory. In the second order approximation the
matrix element $V_{21}$ for the transition between states 1 and 2 is written
in the following form \cite{LLQE}
\begin{equation}
 V_{21} = \sum_{n} \Bigl( \frac{V^{\prime}_{2n} V_{n1}}{{\cal E}_1 -
 {\cal E}^{I}_n} + \frac{V_{2n} V^{\prime}_{n1}}{{\cal E}_1 -
 {\cal E}^{II}_n} \Bigr) \label{eq42}
\end{equation}
where the notation ${\cal E}$ designates the total energy of the system
(`molecule + photons'), i.e. in the case considered here we have ${\cal
E}^{I}_n = E_n$ and ${\cal E}^{II}_n = E_n + \omega + \omega^{\prime}$. The
matrix elements $V_{ab}$ represent absorbtion of the photon with the wave
vector ${\bf k}$. Analogously, the matrix elements $V^{\prime}_{ab}$
represent emission of the photon with the wave vector ${\bf k}^{\prime}$. In
the general case, in Eq.(\ref{eq42}) the $V = V_{21}$ interaction is
represented in the form $V = V^{e} + V^{m} + V^{qe} + V^{qm} + \ldots$ and
$V^{\prime} = (V^{e})^{\prime} + (V^{m})^{\prime} + (V^{qe})^{\prime} +
(V^{qm})^{\prime} + \ldots$, where $V^{e}, V^{m}, V^{qe}$ are the electric
dipole, magnetic dipole and electric quadruple interactions, respectively.
Keeping only lowest order terms in the expansion of $V$ in terms of the
fine-structure constant $\alpha \approx \frac{1}{137}$, we can write $V
\approx V^{e} + V^{m}$ and $V^{\prime} \approx (V^{e})^{\prime} +
(V^{m})^{\prime}$. In this case one finds from Eq.(\ref{eq42})
\begin{eqnarray}
 V_{21} = \sum_{n} \Bigl[ \frac{(V^{e})^{\prime}_{2n} V^{e}_{n1}}{{\cal
 E}_1 - {\cal E}^{I}_n} +
 \frac{(V^{e})_{2n} (V^{e})^{\prime}_{n1}}{{\cal E}_1 - {\cal E}^{II}_n}
 \Bigr]
 + \sum_{n} \Bigl[ \frac{(V^{e})^{\prime}_{2n} V^{m}_{n1}}{{\cal
 E}_1 - {\cal E}^{I}_n} +
 \frac{(V^{e})_{2n} (V^{m})^{\prime}_{n1}}{{\cal E}_1 - {\cal E}^{II}_n}
 + \frac{(V^{m})^{\prime}_{2n} V^{e}_{n1}}{{\cal
 E}_1 - {\cal E}^{I}_n} \label{equa43} \\
 + \frac{(V^{m})_{2n} (V^{e})^{\prime}_{n1}}{{\cal E}_1 - {\cal E}^{II}_n}
 \Bigr] + \sum_{n} \Bigl[ \frac{(V^{m})^{\prime}_{2n} V^{m}_{n1}}{{\cal
 E}_1 - {\cal E}^{I}_n} +
 \frac{(V^{m})_{2n} (V^{m})^{\prime}_{n1}}{{\cal E}_1 - {\cal E}^{II}_n}
 \Bigr] + \ldots \nonumber
\end{eqnarray}
By neglecting here by all terms $\sim V^{m} V^{m}$ and other terms of higher
orders in the fine structure constant $\alpha$, we obtain the following
formula for the differential cross-section of light scattering $d \sigma$
\begin{equation}
 d\sigma = \mid V_{21} \mid^2 \frac{(\omega^{\prime})^2 do^{\prime}}{4
 \pi^2} = d\sigma_{ee} + d\sigma_{em} \label{equ44}
\end{equation}
where $d\sigma_{ee}$ is the part of the total cross-section which can be
reduced to the expression given above (see Eq.(\ref{eq35})). This part of
the cross-section is not related with the optical activity. The second term
in the right-hand side of Eq.(\ref{equ44}) is significantly smaller, in the
general case, than the first term, i.e. $d\sigma_{em} \ll d\sigma_{ee}$.
However, the second term in Eq.(\ref{equ44}) is a great interest, since it
represents all lowest order effects which are determined by the molecular
optical activity.

As follows from Eq.(\ref{equa43}) in order to determine the part of the
total cross-section responsible for molecular optical activity in the lowest
order approximation we need to obtain the explicit formulas for the matrix
elements of the $V^{e} V^{e}, V^{e} V^{m},$ and $V^{m} V^{e}$ products. The
arising expressions are extremely complicated, since each of the $V^{e}$
and/or $V^{m}$ interactions contains an infinite number of $V^{e}_{{\bf e}
\omega}$ and $V^{m}_{{\bf e} \omega}$ components. In the $V^{e} V^{m}$
and/or $V^{m} V^{e}$ products one finds an infinite number of cross-terms
which explicitly depend upon coordinates. These terms cannot be computed
without a complete and accurate knowledge of the molecular electron density
$\rho_e({\bf r})$.

However, we can introduce an approximation that the wavelengths $\lambda$ of
the incident and final photons are significantly larger than typical linear
sizes of molecule $a$ (our light scatterer). In this case we have ${\bf k}
\cdot {\bf r} \le \mid {\bf k} \mid \mid {\bf r} \mid \ll \frac{a}{\lambda}
\approx 0$. In this approximation one finds from Eqs.(\ref{eq40}) and
(\ref{eq41})
\begin{equation}
 {\bf E}_{{\bf e} \omega} = -\imath \sqrt{2 \pi \omega} {\bf e}
 exp(-\imath \omega t)
\end{equation}
and
\begin{equation}
 {\bf H}_{{\bf e} \omega} = \imath \sqrt{2 \pi \omega} ({\bf n} \times
 {\bf e}) exp(-\imath \omega t) \; \; \; .
\end{equation}
Therefore, we can write
\begin{equation}
 V^{e}_{{\bf e} \omega} = \imath \sqrt{2 \pi \omega} ({\bf d} \cdot {\bf
 e}) exp(-\imath \omega t) = \imath \sqrt{2 \pi \omega} ({\bf
 d}_{\omega} \cdot {\bf e})
\end{equation}
and
\begin{equation}
 V^{m}_{{\bf e} \omega} = -\imath \sqrt{2 \pi \omega} [{\bf m} \cdot
 ({\bf n} \times {\bf e})] exp(-\imath \omega t) = -\imath \sqrt{2 \pi
 \omega} [{\bf m}_{\omega} \cdot ({\bf n} \times {\bf e})]
\end{equation}
where ${\bf d}_{\omega}$ and ${\bf m}_{\omega}$ are the corresponding
Fourier-components of the dipole and magnetic moments of the molecule. Note
that with the identity ${\bf m} \cdot ({\bf n} \times {\bf e}) = {\bf e}
\cdot ({\bf m} \times {\bf n})$, the formula for magnetic interaction can
also be written in another form
\begin{equation}
 V^{m}_{{\bf e} \omega} = -\imath \sqrt{2 \pi \omega} [({\bf m}_{\omega}
 \times {\bf n}) \cdot {\bf e}]
\end{equation}
which is similar to the formula for $V^{e}_{{\bf e} \omega}$ in which the
dipole moment ${\bf d}$ is replaced by the vector-product ${\bf m}_{\omega}
\times {\bf n}$.

Now, we can write the lowest order term upon the magnetic interaction in the
differential cross-section $d\sigma$ of the light scattering
\begin{eqnarray}
 d\sigma = \Big| \sum_{n} \frac{({\bf d}_{2n} \cdot {\bf e}^{\prime})
 ({\bf d}_{n1} \cdot {\bf e})}{\omega_{n1} - \omega - \imath 0} +
 \frac{({\bf d}_{2n} \cdot {\bf e}) ({\bf d}_{n1} \cdot {\bf
 e}^{\prime})}{\omega_{n1} + \omega^{\prime} - \imath 0} \Bigr|
 \Big| \sum_{n} \frac{({\bf d}_{2n} \cdot {\bf e}^{\prime})
 [({\bf m}_{n1} \times {\bf n}) \cdot {\bf e}]}{\omega_{n1} - \omega -
 \imath 0} + \\
 \frac{[({\bf m}^{*}_{2n} \times {\bf n}) \cdot {\bf e}^{\prime}] ({\bf
 d}_{n1} \cdot {\bf e})}{\omega_{n1} - \omega - \imath 0} +
 \frac{({\bf d}_{2n} \cdot {\bf e})
 [({\bf m}^{*}_{n1} \times {\bf n}) \cdot {\bf e}^{\prime}]}{\omega_{n1} +
 \omega^{\prime} - \imath 0}
 + \frac{[({\bf m}_{2n} \times {\bf n}) \cdot {\bf e}] ({\bf d}_{n1} \cdot
 {\bf e}^{\prime})}{\omega_{n1} + \omega^{\prime} - \imath 0} \Bigr| \cdot
 \frac{\omega (\omega^{\prime})^3}{\hbar^2 c^4} do^{\prime} \nonumber
\end{eqnarray}
where the notation ${\bf e}^{\prime}$ designates the vector $({\bf
e}^{\prime})^{*}$. This notation is also used in the two following
equations. This equation can be re-written as
\begin{eqnarray}
 d\sigma = \Big| \sum_{n} \frac{({\bf d}_{2n} \cdot {\bf e}^{\prime})
 ({\bf d}_{n1} \cdot {\bf e})}{\omega_{n1} - \omega - \imath 0} +
 \frac{({\bf d}_{2n} \cdot {\bf e}) ({\bf d}_{n1} \cdot {\bf
 e}^{\prime})}{\omega_{n1} + \omega^{\prime} - \imath 0} \Bigr|
 \Big| \sum_{n} \frac{({\bf d}_{2n} \cdot {\bf e}^{\prime})
 [{\bf m}_{n1} \cdot ({\bf n} \times {\bf e})]}{\omega_{n1} - \omega -
 \imath 0} + \label{cross1} \\
 \frac{[{\bf m}^{*}_{2n} \cdot ({\bf n} \times {\bf e}^{\prime})] ({\bf
 d}_{n1} \cdot {\bf e})}{\omega_{n1} - \omega - \imath 0} +
 \frac{({\bf d}_{2n} \cdot {\bf e})
 [{\bf m}^{*}_{n1} \cdot ({\bf n} \times {\bf e}^{\prime})]}{\omega_{n1} +
 \omega^{\prime} - \imath 0}
 + \frac{[{\bf m}_{2n} \cdot ({\bf n} \times {\bf e})] ({\bf d}_{n1} \cdot
 {\bf e}^{\prime})}{\omega_{n1} + \omega^{\prime} - \imath 0} \Bigr| \cdot
 \frac{\omega (\omega^{\prime})^3}{\hbar^2 c^4} do^{\prime} \nonumber
\end{eqnarray}
In these equations and below the notation ${\bf m}^{*}$ stands for the
vector which is a complex conjugate vector to the vector of magnetic dipole
moment ${\bf m}$. In quantum mechanics (in the coordinate representation)
we always have ${\bf d}^{*} = {\bf d}$, but ${\bf m}^{*} \ne {\bf m}$. Note
that the vector ${\bf n}$ in these equations corresponds to the direction of
the scattered light. Formally, this vector can be oriented in an arbitrary
spatial direction, but in almost all modern experiments on optical activity
in homogeneous solutions the direction of the scattered light always
coincides with the direction of the incident light. This means that our
differential cross-section must be multiplied by a delta-function
$\delta({\bf n}_{in} - {\bf n})$ and integrated over the angular variables
$o^{\prime} = (\theta^{\prime}, \phi^{\prime})$ of the unit vector ${\bf n}
= (cos\theta^{\prime} cos\phi^{\prime}, cos\theta^{\prime} sin\phi^{\prime},
sin\theta^{\prime})$ which represents the direction of the final photon. The
unit vector ${\bf n}_{in}$ describes the direction of the incident photon.
This produces the following expression for the cross-section $\sigma$
\begin{eqnarray}
 \sigma = \frac{4 \pi \omega (\omega + \omega_{12})^3}{\hbar^2 c^4} \cdot
 \Big| \sum_{n} \frac{({\bf d}_{2n} \cdot {\bf e}^{\prime})
 ({\bf d}_{n1} \cdot {\bf e})}{\omega_{n1} - \omega - \imath 0} +
 \frac{({\bf d}_{2n} \cdot {\bf e}) ({\bf d}_{n1} \cdot {\bf
 e}^{\prime})}{\omega_{n1} + \omega^{\prime} - \imath 0} \Bigr|
 \Big| \sum_{n} \frac{({\bf d}_{2n} \cdot {\bf e}^{\prime})
 [{\bf m}_{n1} \cdot ({\bf n}_{in} \times {\bf e})]}{\omega_{n1} - \omega
 - \imath 0} \nonumber \\
 + \frac{[{\bf m}^{*}_{2n} \cdot ({\bf n}_{in} \times {\bf e}^{\prime})]
 ({\bf d}_{n1} \cdot {\bf e})}{\omega_{n1} - \omega - \imath 0} +
 \frac{({\bf d}_{2n} \cdot {\bf e})
 [{\bf m}^{*}_{n1} \cdot ({\bf n}_{in} \times
 {\bf e}^{\prime})]}{\omega_{n1} + \omega^{\prime} - \imath 0}
 + \frac{[{\bf m}_{2n} \cdot ({\bf n}_{in} \times {\bf e})] ({\bf d}_{n1}
 \cdot {\bf e}^{\prime})}{\omega_{n1} + \omega^{\prime} - \imath 0} \Bigr|
 \label{cross3}
\end{eqnarray}
where $\omega^{\prime} = \omega + \omega_{12}$ and unit-vector ${\bf
n}_{in}$ designates the direction of propagation of the incident photon.

The expression, Eq.(\ref{cross3}), can be cast in the following form
\begin{eqnarray}
 \sigma = \frac{4 \pi \omega (\omega + \omega_{12})^3}{\hbar^2 c^4} \cdot
 \Bigl| (C_{ik})_{21} ({\bf e}^{\prime}_i)^{*} {\bf e}_k \Bigr| \cdot
 \Bigl| (\hat{S}_{ik})_{21} ({\bf e}^{\prime})^{*}_i ({\bf n}_{in} \times
 {\bf e})_k +
 (\hat{T}_{ik})_{21} ({\bf n}_{in} \times ({\bf e}^{\prime})^{*})_i
 {\bf e}_k \\
+ (\hat{U}_{ik})_{21} ({\bf e})_i ({\bf n}_{in} \times ({\bf
 e}^{\prime})^{*})_k + (\hat{V}_{ik})_{21} ({\bf n}_{in} \times {\bf
 e}^{\prime})_i ({\bf e}^{\prime}_k)^{*} \Bigr| \nonumber
\end{eqnarray}
where $(S_{ik})_{21}, (T_{ik})_{21}, (U_{ik})_{21}$ and $(V_{ik})_{21}$ are
$3 \times 3$ tensors, while the dipole-dipole tensor $(C_{ik})_{21}$ is
defined above in Eq.(\ref{eq36}). Here we assume that, in the general case,
the vectors ${\bf e}^{\prime}$ and ${\bf e}$ which represent the
polarization of light are complex. Each of these tensors is represented as a
sum of its irreducible components, e.g., $S_{ik} = S^{0} \delta_{ik} +
S^{s}_{ik} + S^{a}_{ik}$, where
\begin{equation}
 S^{0} = \frac13 S_{ii} \; \; \; , \; \; \;
 S^{s}_{ik} = \frac12 (S_{ik} + S_{ki}) - S^{0} \delta_{ik} \; \; \; , \; \;
 \; S^{a}_{ik} = \frac12 (S_{ik} - S_{ki})
\end{equation}
Note also that $S^{0},T^{0}, U^{0}$ and $V^{0}$ are called the scalar parts
of the $S, T, U$ and $V$ tensors, respectively. The components with the
superscripts $s$ and/or $a$ (e.g., $T^{s}, T^{a}$) are the symmetric and
antisymmetric parts of the tensor. All components of the $S^{0}, T^{0},
U^{0}, V^{0}, S^{s}_{ik}, T^{s}_{ik}, U^{s}_{ik}, V^{s}_{ik}, S^{a}_{ik},
T^{a}_{ik}, U^{a}_{ik}$ and $V^{a}_{ik}$ tensors contain the products of the
corresponding components of the ${\bf d}$ and ${\bf m}$ vectors, which are
the vectors of the electric dipole moment and magnetic dipole moment,
respectively. The vector-operators which represent the electric and magnetic
dipole momenta are assumed to be self-conjugate. Furthermore, as mentioned
above in the coordinate representation the vector ${\bf d}$ is a real vector
(i.e. ${\bf d}^{*} = {\bf d}$), while the vector ${\bf m}$ is a complex
vector (i.e. ${\bf m}^{*} \ne {\bf m}$). For instance, the explicit
expressions for the $S^{0}, T^{0}, U^{0}$ and $V^{0}$ tensors (they are
called the scalar-components) are
\begin{eqnarray}
 (S^{0})_{21} = \frac13 \sum_{n} \frac{(d_i)_{2n}
 (m_i)_{n1}}{\omega_{n1} - \omega} \; \; \; , \; \; \; \;
 (T^{0})_{21} = \frac13 \sum_{n} \frac{(m^{*}_i)_{2n}
 (d_i)_{n1}}{\omega_{n1} - \omega} \; \; \; , \label{tens} \\
 (U^{0})_{21} = \frac13 \sum_{n} \frac{(d_i)_{2n}
 (m^{*}_i)_{n1}}{\omega_{n2} + \omega} \; \; \; , \; \; \; \;
 (V^{0})_{21} = \frac13 \sum_{n} \frac{(m_i)_{2n}
 (d_i)_{n1}}{\omega_{n2} + \omega} \; \; \; , \nonumber
\end{eqnarray}
respectively. Analogous formulas for the symmetric and antisymmetric parts
of the $S, T, U$ and $V$ tensors are significantly more complicated. These
formulas and the physical meaning of all irreducible components of these $S,
T, U$ and $V$ tensors will be discussed elsewhere.

Thus, we have shown that all phenomena related to the optical activity can
completely be described with the use of only four tensors: $\hat{S}_{21},
\hat{T}_{21}, \hat{U}_{21}$ and $\hat{V}_{21}$. The fifth tensor
$\hat{C}_{21}$ (the tensor of electric-dipole light scattering) is included
in the formula for the cross-section as an amplification factor. These five
tensors have fifteen irreducible tensor-components $C^{0}, C^{s}_{ik},
C^{a}_{ik}, S^{0}, T^{0}, U^{0}, V^{0}, S^{s}_{ik}, T^{s}_{ik}, U^{s}_{ik},
V^{s}_{ik}, S^{a}_{ik}, T^{a}_{ik}, U^{a}_{ik}$ and $V^{a}_{ik}$. The first
three tensors $C^{0}, C^{s}_{ik}, C^{a}_{ik}$ here have nothing to do with
the optical activity itself. Instead they determine the amplification factor
which also appears to be $\omega-$dependent. The optical activity is
described by the twelve tensors ($S^{0}, T^{0}, U^{0}, V^{0}, S^{s}_{ik},
T^{s}_{ik}, U^{s}_{ik}, V^{s}_{ik}, S^{a}_{ik}, T^{a}_{ik}, U^{a}_{ik}$ and
$V^{a}_{ik}$). In many real applications, however, the total number of
independent tensors can be reduced. For instance, if the $1$- and $2$-states
are identical and $\omega_{21} = 0$ (Rayleigh scattering), then to describe
optical activity one needs only two tensors (not four!) with six irreducible
components. This case corresponds to the regular optical activity (optical
rotation) measured in modern experiments with dilute solutions of organic
molecules. Furthermore, if the polarization vectors are always chosen as
real (not complex), then to describe the optical activity one needs only one
$3 \times 3$ tensor with three irreducible components. However, the explicit
$\omega-$dependence of such a tensor will be quite complicated. All such
cases will be considered in our next study.

The intensity of the scattered light $I^{\prime}$ is uniformly related to
the intensity of the incident light $I$ by the relation
\begin{equation}
  I^{\prime} = \Bigl(\frac{\omega^{\prime}}{\omega}\Bigr) \sigma I
\end{equation}
As follows from the formula for the cross-section $\sigma$,
Eq.(\ref{cross3}), in any optically active solution the intensity of the
(scattered) light will always be rotated during its propagation along the
direction ${\bf n}_{in}$. The factor
$\Bigl(\frac{\omega^{\prime}}{\omega}\Bigr) \sigma$ in the last formula can
be considered as the rotation power. As follows from the last formula the
uniform combination of the twelve tensors mentioned above multiplied by the
amplification factor, Eq.(\ref{eq36}), allows one to determine the so-called
rotation power of the given optically active solution for the initially
polarized light. Note that only our approach produces the correct and
complete formula for the $\omega-$dependence of the rotation power.

\section{Quantum theory of molecular optical activity}

As can be seen above, the physical origin of the relations between different
parameters used in classical theory of optical activity remains unknown.
The corresponding analytical expressions, numerical values and all possible
relations between such `phenomenological' parameters can be found only with
the use of modern quantum theory based on Quantum Electrodynamics. The first
and very important step in the development of quantum origin of optical
activity was made by Rosenfeld almost 80 years ago \cite{Rose}. A simplified
and complete version of the approach developed by Rosenfeld can be found,
e.g., in \cite{EWC}. Below, we shall follow the same general direction. Our
main goal in this Section is to obtain the relation between the optical
rotatory parameter $\beta$ from Eq.(\ref{Ebas1}) and properties of an
isolated molecule. For an isolated molecule the optical rotatory parameter
$\beta$ can only be a function of the molecular $2^{\ell}$-pole moments. In
reality, however, only a few moments with small $\ell$ ($\ell = 1, 2$)
contribute noticeably. As we mentioned above the parameter $\beta$ is a
pseudoscalar. Therefore, the first (largest) term in the expansion of
$\beta$ in terms of $2^{\ell}$-pole molecular moments is proportional to the
scalar product of the dipole vector and the pseudovector of the magnetic
moment ${\bf d} \cdot {\bf m}$. The second term must be proportional to the
product ${\bf d} \cdot \hat{Q} \cdot {\bf m}$, where $\hat{Q}$ is the second
order tensor of the electric quadropole moment.

Below, we shall assume that all molecular wave functions (for the ground and
excited states) are known (or can be determined) to very good accuracy. In
this case, by using Rosenfeld's formula one can calculate the optical
rotatory parameter $\beta$ (in some studies it is also called the chiral
response parameter)
\begin{eqnarray}
 \beta = \frac{c}{6 \pi \hbar} \sum_b \frac{Im\Bigl[ \langle a \mid {\bf d}
 \mid b \rangle \langle b \mid {\bf m} \mid a \rangle\Bigr]}{\nu^2_{ab} -
 \nu^2}
\end{eqnarray}
where the summation is taken over all intermediate states. In this equation
we use the linear frequencies $\nu$ instead of circular frequencies
$\omega$, where $\omega = 2 \pi \nu$. The notation $Im$ designates the
imaginary part of the terms written in brackets. Symbols $a$ and $b$ stand
for the quantum (molecular) states, while $\mid a \rangle$ and $\mid b
\rangle$ mean the corresponding wave functions. Rosenfeld's formula is based
on an assumption that all molecular states (ground and excited) have zero
widths. In other words, these states are stable, i.e. the decay time is
infinite. In general, this is not a very realistic assumption and we need to
introduce finite line widths, e.g., $\gamma_{ab}(\nu) = \frac{4 \pi^2 e^2
|{\bf D}_{ab}|^2 (\nu_a - \nu_b)^3}{3 \epsilon_0 \hbar c^3}$ (in SI-units
and in the lowest order dipole approximation \cite{Loud}). The Rosenfeld
formula for the optical rotation $\delta$ can now be written in the form
\begin{equation}
 \delta = \frac{16 \pi^2 N z}{3 h c} \sum_b \frac{\nu^2 R_{ab}}{\nu^2_{ab}
 - \nu^2 + \imath \nu \gamma_{ab}}
\end{equation}
where $R_{ab} = Im\Bigl[ \langle a \mid {\bf d} \mid b \rangle \langle b
\mid {\bf m} \mid a \rangle\Bigr]$ is the so-called rotating power. By
separating the real and imaginary parts of this expression one finds for the
actual optical rotation
\begin{eqnarray}
 \theta = \frac{16 \pi^2 N z}{3 h c} \sum_b \frac{\nu^2 (\nu^2_{ab} -
 \nu^2) R_{ab}}{(\nu^2_{ab} - \nu^2)^2 + \nu^2 \gamma^2_{ab}} \label{rot}
\end{eqnarray}
and for the circular dichroism
\begin{eqnarray}
 \kappa = - \frac{16 \pi^2 N z}{3 h c} \sum_b \frac{\nu^3 \gamma_{ab}
 R_{ab}}{(\nu^2_{ab} - \nu^2)^2 + \nu^2 \gamma^2_{ab}} \label{dich}
\end{eqnarray}
where the notations from formula Eq.(\ref{KK5}) are used. The definition of
rotating power given above corresponds to the dipole-dipole approximation.
In the higher order approximation the rotating power must be taken in the
form $R_{ab} = Im\Bigl[ \langle a \mid {\bf d} \mid b \rangle \langle b \mid
{\bf m} \mid a \rangle + \sum_{cd} \langle a \mid {\bf d} \mid c \rangle
\langle c \mid \hat{Q} \mid d \rangle \langle d \mid {\bf m} \mid a
\rangle\Bigr]$. In reality it is very difficult to calculate the matrix
elements $R_{ab}$ accurately. However, a number of useful approximate
formulas have been derived from the expressions Eq.(\ref{rot}) and
Eq.(\ref{dich}). For instance, if in some experiment we can see $N$ peaks in
the $\theta(\lambda)$ function and $K$ peaks in the $\kappa(\lambda)$
function, then it is possible to approximate our experimental data by using
the two following formulas
\begin{eqnarray}
 \theta(\lambda) = \sum^N_{i=1} \frac{A_i (\lambda^2 -
 \lambda_{i}^{2})}{(\lambda^2 - \lambda_{i}^{2})^2 + B_i} \; \; \; and
 \; \; \; \kappa(\lambda) = \sum^K_{j=1} \frac{C_j \lambda}{(\lambda^2 -
 \lambda_{j}^{2})^2 + D_j}
\end{eqnarray}
where all numerical parameters $A_i, B_i, C_i$ and $D_i$ must be determined
by using the experimental values of $\theta$ and $\kappa$ at different
wavelengths $\lambda = \frac{1}{\nu}$. Numerical examples can be found in
the book by Djerassi \cite{Djer}.

Rosenfeld's theory of optical activity allows one to determine the relations
between basic molecular properties and actual optical rotation and circular
dichroism observed in experiments. Indeed, by applying the known molecular
wave functions one can compute the values of $\langle a \mid {\bf d} \mid b
\rangle$ and $\langle b \mid {\bf m} \mid a \rangle$ which are used in
formulas for $\theta$ and $\kappa$ above. By using these values we can
evaluate the rotating powers $R_{ab}$. Then we can try to approximate the
curves $\theta(\nu)$ and $\kappa(\nu)$ obtained in actual experiments.
During this step all line widths $\gamma_{ab}$ can be varied as numerical
parameters. In practice, this approach works approximately only for some
simple molecules. For many molecules of interest, e.g., for complex
molecules used in cancer research, the current accuracy of the numerical
determination of the $\langle b \mid {\bf m} \mid a \rangle$ values is not
sufficient to make accurate comparisons with experiments. In particular, the
signs of the $\langle b \mid {\bf m} \mid a \rangle$ values can be wrong in
a number of cases. In addition to this, Rosenfeld's theory of optical
activity is essentially a semi-classical theory, since all radiation fields
in this theory are considered classically. The most rigorous analysis of
molecular optical activity can be performed only on a basis of modern
quantum electrodynamics (QED) \cite{LLQE}. This will be one of our goals in
future studies.

\section{Specific rotation by chiral organic molecules}

In previous Sections we have briefly considered some theoretical aspects of
the molecular optical activity at arbitrary wavelengths. In this Section we
discuss a few basic features which are known for optically active organic
molecules in solutions. In general, if some organic molecule has non-zero
electric and magnetic moments, then its optical rotation $\theta$,
Eq.(\ref{rot}), differs from zero. Such a molecule shows a number of
phenomena which are usually designated as `optical activity', or briefly, as
an optical rotation of plane-polarized light. In principle, any molecule
which does not coincide with its mirror image can be optically active.
Moreover, for each optically active molecule one can always find another
non-identical form of the same molecule which is related to the original
molecule through reflection. These two forms of one molecule are called
enantiomers, specifically $D-$ and $L-$ enantiomers, which naturally refer
to the right- and left-handed forms, respectively. In many sources the D-
and L-enantiomers of various molecules are discussed. However, it should be
emphasized that currently there is no uniform relation between the absolute
configuration of complex molecules and their ability to be left- and/or
right-rotating. Here it is not our intention to summarize all basic rules
found for numerous organic substances which are optically active. Instead,
we restrict ourselves to an analysis relevant to observation of optical
activity in the vacuum ultraviolet region.

The most interesting cases can be observed in various organic molecules,
i.e. in molecules which include one or more carbon atoms. Formally, one
carbon atom in a molecule which is bonded to four different atoms and/or
groups of atoms is sufficient for manifestation of optical activity. In
general, such a carbon atom is called an asymmetric atom, or a chiral
center. In many cases the optical activity can be observed in molecules with
two, three and more asymmetric carbon atoms (for more detail, see, e.g.,
\cite{Mason}, \cite{Rober} and references therein). A very well known
example is the tartaric acid which may exist in the form of D- and
L-enantiomers and in its meso-form which has no optical activity. Note also
that a number of organic molecules with no chiral centers show overall
optical activity, e.g., allenes, spiranes and biphenyls. Such systems are
considered as inherently dysymmetric. The active electrons in these
molecules are delocalized over a chiral nuclear system.

In this work we restrict ourselves to the consideration of organic molecules
with one asymmetric carbon atom (chiral center). In general, the observed
optical rotation $\theta$, Eq.(\ref{rot}), produced by one asymmetric carbon
atom in a molecule will be small. However, if some additional conditions are
combined with each other, then the actual optical rotation $\theta$
increases to moderate, large and very large values. It was shown more than
sixty years ago that `close' presence of some special groups of atoms can
increase the actual optical rotation by a few orders of magnitude. Such
special groups are called `chromophors'. Typical examples of chromophors
are the $-$NH$_2$, $>$CO, $-$CN, $-$C$_6$H$_5$ groups and some others. It
should be emphasized that none of these groups is optically active, but each
amplifies significantly the optical activity of the neithbouring chiral
center.

Currently, there are a few dozens of different atomic groups which are
recognized as regular basic chromophors and a large number of special groups
of atoms which become chromophors only at certain wavelengths. In general,
any group of atoms which has excessive $\pi-$electron density can be
considered as a potential chromophor and any experimental study of optical
activity in organic molecules is reduced to the analysis of various
chromophors and their influence on one of more asymmetric carbon atom(s).
The problem contains many complications. For instance, if one of the
hydrogen atoms in a benzene ring bonded with a chiral carbon center is
replaced by the $-$NH$_2$ group, then one finds the new chromophor
$-$C$_6$H$_4$NH$_2$ which has a different influence on this chiral center.
In other words, the change in optical activity of the chiral
center produced by the new chromophor ($-$C$_6$H$_4$NH$_2$) cannot be
predicted accurately and uniformly from the analogous information known for
the $-$C$_6$H$_5$ and $-$NH$_2$ groups. In general, the delocalization of
$\pi-$electrons in various organic molecules can be used to create a huge
number of `new' chromophors. Moreover, if the same chiral center is bonded
to both the $-$C$_6$H$_5$ and $-$NH$_2$ groups, the amplification of its
chiral activity will be drastically different from the previous case. In
addition to this one also finds that substitution of one hydrogen atom in
the $-$C$_6$H$_5$ chromophor by, e.g., the $-$CH$_3$ group will also have a
noticeable effect on overall optical activity. The following experimental
analysis must detect (and investigate) a direct relation between the actual
optical activity and position of the hydrogen atom (in the benzene ring)
which was replaced by the $-$CH$_3$ group.

It is important for the general theory that each chromophore group can be
represented by a number of poles located in the complex plane of frequencies
$\nu$ (so-called $\nu$-plane). In other words, any chromophore group has a
number of poles in the complex $\nu$-plane associated with it. If such a
pole is close to the real axis, then in experiments one finds large and very
large values of optical rotation $\delta$ and circular dichroism $\kappa$,
respectively. In real organic molecules one finds not one, but a few
different chromophore groups. The experimental curves for the $\delta(\nu)$
and $\kappa(\nu)$ values measured for $\nu > 0$ in actual organic molecules
are the result of interaction between various poles located in a complex
frequency plane ($\nu$) at different frequencies. In general, the
interaction between different chromophores produces very complicated spectra
for the optical rotation $\delta(\nu)$ and circular dichroism $\kappa(\nu)$.
The complexity of these spectra rapidly increases as the number of poles per
unit frequency interval increases. In particular, this is the case for
vacuum ultraviolet wavelengths, since almost all known chromophors have many
absorbtion lines located in VUV region.

\section{Conclusion}

We have considered the phenomena of optical activity in homogeneous
solutions of various organic substances. The classical macroscopic theory
based on Maxwell equations in dielectric (or nonconducting) media is
discussed in detail. The Stokes parameters for almost monochromatic light
are defined rigorously. The relations between the optical rotation and
circular dichroism are derived from the basic Kramers-Kronig relations.
These relations allow one to obtain/evaluate, e.g., the circular dichroism
by using the known values of optical rotation at the same frequencies. The
explicit expression for the tensors of molecular optical activity are
derived. Our formulas derived for the tensor(s) of molecular optical
activity can be used to explain a large number of phenomenon currently
known in molecular optical activity. Note that our formulas can
successfully be applied to the case of the Rayleigh (or non-shifted)
scattering when $\omega_{21} = 0$ and also to the cases when $\omega_{21}
\ne 0$ (shifted or Raman light scattering). It is shown that all known
lowest order effects of optical activity must be described with the use of
finite number of tensors (five, three or one tensors).

We also briefly consider the quantum (or semi-classical) theory of molecular
optical activity developed by Rosenfeld in \cite{Rose}. In this theory all
molecules are quantum systems, while all electromagnetic fields are
described by classical Maxwell equations. A possibility to extend
measurements of optical rotation and circular dichroism into the vacuum
ultraviolet region is discussed. Currently, this task seems to be extremely
difficult, since there are a large number of unsolved problems which must be
considered before the whole procedure can be usefully implemented. Moreover,
it is clear that even the requisite experimental technique will have many
fundamental differences from the technique applied for traditional
wavelengths. Nevertheless, we can expect that measurements of optical
rotation and circular dichroism in the vacuum ultraviolet region will
produce a large volume of very valuable experimental data. These
measurements will open a new avenue for some important discoveries and
improvements in our current understanding of optical activity of organic
molecules. In general, such measurements can be performed with the use of
high quality VUV-radiation from modern electron synchrotrons.

It should be mentioned in conclusion that many experimental and theoretical
methods developed earlier to observe/describe the optical activity in
solutions of organic substances are now a great interest for modern Stellar
Astrophysics (see, e.g., \cite{Ital}, \cite{Clark} and references therein).
It was shown that of radiation emitted by many Stellar objects is partially
polarized ($\approx$ 1 - 10 \%). In particular, the polarization of
radiation emitted by the hot Be-stars is discussed in \cite{Jon1} -
\cite{Mof}. We shall not go into discussion of these and other similar
problems in this article. Note only that the overall complexity of
Astrophysical problems related with the polarization of radiation is
significantly higher than for `regular' problems known from Organic
Chemistry.

\begin{center}
   {\bf Acknowledgments}
\end{center}

It is a pleasure to acknowledge the University of Western Ontario for
financial support.

\end{document}